\documentstyle[prl,epsf,aps]{revtex}

\newtheorem{theorem}{Theorem}[section]

\newtheorem{remark}{Remark}

\newtheorem{lemma}[theorem]{Lemma}
\newtheorem{proposition}{Proposition}
\newtheorem{corollary}{Corollary}
\newtheorem{conjecture}{Conjecture}
\newtheorem{assumption}{Assumption}

\newcommand{\bpr} {\noindent {\bf Proof }}
\newcommand{\epr} {$\Box$\\}
\newcommand{\bd}{
\begin{document}}
\newcommand{\ed}{\end{document}}
\newcommand{\beq}{\begin{equation}}
\newcommand{\eeq}{\end{equation}}
\newcommand{\bef}{\begin{figure}}
\newcommand{\enf}{\end{figure}}
\newcommand{\bea}{\begin{eqnarray}}
\newcommand{\eea}{\end{eqnarray}}
\newcommand{\bth}{\begin{theorem}}
\newcommand{\eth}{\end{theorem}}
\newcommand{\bhyp}{\begin{assumption}}
\newcommand{\ehyp}{\end{assumption}}
\newcommand{\bp}{\begin{proposition}}
\newcommand{\ep}{\end{proposition}}
\newcommand{\bco}{\begin{corollary}}
\newcommand{\eco}{\end{corollary}}
\newcommand{\bconj}{\begin{conjecture}}
\newcommand{\econj}{\end{conjecture}}
\newcommand{\ble}{\begin{lemma}}
\newcommand{\ele}{\end{lemma}}
\newcommand{\bR}{\begin{remark}}
\newcommand{\eR}{\end{remark}}
\newcommand{\bc}{\begin{center}}
\newcommand{\ec}{\end{center}}
\newcommand{\ben}{\begin{enumerate}}
\newcommand{\een}{\end{enumerate}}
\newcommand{\bit}{\begin{itemize}}
\newcommand{\eit}{\end{itemize}}
\newcommand{\su}{\section}
\newcommand{\ssu}{\subsection}
\newcommand{\sssu}{\subsubsection}
\newcommand{\nid}{\noindent}
\newcommand{\nnb}{\nonumber}

\newcommand\bbbr{{\sf I\!R}}
\newcommand\bbbc{{\sf I\!C}}
\newcommand\bbbn{{\sf I\!N}}
\newcommand\bbbh{{\sf I\!H}}
\newcommand\bbbz{{\sf Z\!\!Z}}

\newcommand\calA{{\cal A}}
\newcommand\calB{{\cal B}}
\newcommand\calC{{\cal C}}
\newcommand\calD{{\cal D}}
\newcommand\calE{{\cal E}}
\newcommand\calF{{\cal F}}
\newcommand\calG{{\cal G}}
\newcommand\calH{{\cal H}}
\newcommand\calI{{\cal I}}
\newcommand\calK{{\cal K}}
\newcommand\calL{{\cal L}}
\newcommand\calM{{\cal M}}
\newcommand\calN{{\cal N}}
\newcommand\calP{{\cal P}}
\newcommand\calQ{{\cal Q}}
\newcommand\calR{{\cal R}}
\newcommand\calS{{\cal S}}
\newcommand\calW{{\cal W}}
\newcommand\calU{{\cal U}}
\newcommand\calV{{\cal V}}
\newcommand\calT{{\cal T}}
\newcommand\calX{{\cal X}}
\newcommand\calZ{{\cal Z}}
\newcommand\call{{\cal l}}
\newcommand\caln{{\cal n}}
\newcommand\calr{{\cal r}}

\newcommand\bE{{\bf E}}
\newcommand\bF{{\bf F}}
\newcommand\bX{{\bf X}}
\newcommand\bY{{\bf Y}}
\newcommand\bU{{\bf U}}
\newcommand\bZ{{\bf Z}}
\newcommand\ba{{\bf a}}
\newcommand\bb{{\bf b}}
\newcommand\be{{\bf e}}
\newcommand\bk{{\bf k}}
\newcommand\bi{{\bf i}}
\newcommand\bl{{\bf l}}
\newcommand\bn{{\bf n}}
\newcommand\bq{{\bf q}}
\newcommand\bv{{\bf v}}
\newcommand\bx{{\bf x}}
\newcommand\by{{\bf y}}
\newcommand\bz{{\bf z}}
\newcommand\bw{{\bf w}}

\newcommand{\deq}{\stackrel {\rm def}{=}}

\newcommand{\sep}{\; \,}
\newcommand{\D}{\displaystyle}
\newcommand{\T}{\textstyle}
\newcommand{\etc}{etc $\dots$}
\newcommand{\etal}{etc $\dots$}
\newcommand\dL{^\partial \Lambda}

\def\Appendix{\section*{APPENDIX}}

\baselineskip18pt

\bd

\baselineskip18pt

\title {What can one learn about Self-Organized Criticality
from Dynamical Systems theory ?}

\author {Ph. Blanchard
\thanks{University of Bielefeld, BiBoS, Postfach 100131, D-33501,
Bielefeld, Germany}
B. Cessac
\thanks{Institut Non Lin\'eaire de Nice, 1361 Route des
Lucioles, 06500 Valbonne, France}
T. Kr\"uger
\thanks{University of Bielefeld, BiBoS, Postfach 100131, D-33501,
Bielefeld, Germany and Technische Universitaet, Str. des 17 July 135, 10623, Berlin, Germany}
}
\maketitle

\begin{abstract}
We develop a dynamical system approach for the Zhang's
model of Self-Organized Criticality, for which the
dynamics can be described either 
in terms of Iterated Function Systems, or as a piecewise hyperbolic 
dynamical system of skew-product type. In this setting we describe the SOC
attractor, and discuss its fractal
structure.
We show how  the Lyapunov exponents, the Hausdorff dimensions,
and the system size are related to
the probability distribution of the avalanche size, via  
the Ledrappier-Young formula \cite{LY}.

\end{abstract}

\bigskip

\bigskip

\bigskip

\bigskip

\su{Introduction.}

Within the last 10 years the notion of Self-Organized Criticality (SOC)  became a new paradigm for the 
explanation of a huge variety of phenomena in nature and social sciences. It's origin lies in the 
attemp to explain the widespread appearence of power-law like statistics for characteristic events in a multitude 
of examples like the distribution of the size of earthquakes, 1/f-noise, amplitudes of solar flares,
species extinction .... to name only a very few cases \cite{Bak1,Bak2,Bak3,Jensen}. As a result, an important literature in physics
has been devoted to the study of systems exhibiting SOC.

The complexity of the dynamics in the above mentionned systems is mainly due to the presence of 
long-range spatial and time correlations, leading to non trivial effects like anomalous 
diffusion. At stationarity,  the average incoming flux of external perturbations is 
simply compensated 
by the average outgoing flux that can leave the system at the boundary, or by dissipation in the bulk.
Therefore, there is a constant flux through the system, leading to a non-equilibrium situation. 
What is remarkable in this stationary state,
refered to as the SOC state, is that the distribution of avalanches appears to follow a power 
law, namely there is scale invariance reminiscent of thermodynamic systems at the 
critical point. This is certainly one central reason why SOC has attracted the
physicist community: these systems (apparently) reach {\it spontaneously} a
critical state without  any fine tuning of some control parameter.

Several models have been proposed to mimic these mechanisms like the sandpile model
\cite{Bak1,Bak2}, the abelian sandpile \cite{Dhar} or the continuous energy model 
\cite{Zhang}. Numerical simulations on one hand, and theoretical approaches on the other
hand have lead to a good description of SOC, in particular with respect to the 
computation of critical exponents that are believed to characterize the universality
class the model belongs to, as they do in second order phase transitions.

However, to our knowledge, no serious attemp has been made to study SOC from a dynamical system point
of view
(except \cite{BCK1,Cessac}).
It is however a natural approach
 to try to access the macroscopic behaviour of large sized systems
from the microscopic dynamical evolution.  The macroscopic behaviour at stationarity is characterized 
by a probability measure one has to extract from the microscopic evolution. One is seeking a ``good'' 
measure from a physical point of view, namely a Sinai-Bowen-Ruelle measure (SBR) : in the SOC model
we discuss later this measure maximizes the entropy.  \\

In this paper we develop a dynamical system description for a certain class of SOC models
(like the Zhang's model \cite{Zhang}), for which the whole SOC dynamics can either be described
in terms of Iterated Function Systems, or as a piecewise hyperbolic 
dynamical system of skew-product type 
where  one coordinate encodes the sequence of activations.  Several deep
 results from the theory of hyperbolic dynamical systems can then be used, having interesting
 implications on the SOC dynamics, provided one makes some natural assumption
(like ergodicity) which will be partially justified in this paper. 

With this approach we give a precise definition  of the SOC attractor discussed by some people \cite{Bak1,Bak2}.
We show that it has a fractal structure for low values of the critical energy.
The main objects for which our point of view is appropriate is certainly the 
structure of the asymptotic energy distribution or, in other words, the structure of
the natural invariant measure. We show in particular how  the Lyapunov exponents, the geometric
structure of the support of the invariant measure (Hausdorff dimensions), and the system size are related to
the probability distribution of the avalanche size, via   the Ledrappier-Young formula \cite{LY}.

\su{The dynamical structure of the Zhang model.}

\ssu{Description of the model.}

In this paper we deal
with the Zhang's model on a
$d$ dimensional, connected subgraph $\Lambda \subset \bbbz^d$, with nearest
neighbours edges, though the formalism
we develop holds for more general graphs.
Let $\dL$
be the boundary of $\Lambda$, namely the set of points
in $\bbbz^d/\Lambda$ at distance $1$ from $\Lambda$ and let $N$ the cardinality
 of $\Lambda$.
Each site $i \in \Lambda$
is characterized by its
"energy" $X_i$, which is a non-negative real number. The "state" of the network
is completely defined by the configuration of energies $\bX  = \lbrace
X_i \rbrace_{_{i \in \Lambda}}$.
Let $E_c$ be a real, positive number, called the critical energy,
and
$\calM = [0,E_c[^N$. Let $\D{d_{1,2}(\bX,\bY)}$ be the $L_1$ (resp. $L_2$) distance on $\calM$.  
A configuration
$\bX$ is "stable" iff $\bX \in \calM$
and "unstable" or "overcritical" otherwise. If $\bX$ is stable then
we choose a site $i$ at random with probability $\frac{1}{N}$, and add to it energy
$\delta X$. As far as the physically
 relevant parameter is the local rigidity $\frac{E_c}{\delta X}$ \cite{Piet1},
one can
investigate the cases where $E_c$ varies, and where $\delta X$ is a constant.
 We will therefore assume that $\delta X = 1$
without loss of generality.
If a site $i$ is overcritical ($X_i \geq E_c$), it loses a part of its energy
 in equal parts to its $2d$ neighbours. Namely, we fix a parameter
$\epsilon \in [0,1[$ such that, after relaxation of the site $i$,
the remaining energy of $i$ is $\epsilon X_i$, while the $2d$ neighbours
receive the energy $\frac{(1-\epsilon)X_i}{2d}$.
 Note that in the original
Zhang's model \cite{Zhang}, $\epsilon$ was taken to be zero.
We define here a straightforward extension. Note however that in this paper $\epsilon$ will be considered 
as a  {\it small} parameter
compared to $E_c$.

 If several nodes are simulaneously overcritical,
the local distribution rules are
additively superposed,
i.e. the time evolution of the system is synchronous. The sites of $\dL$ have always
zero energy
(dissipation at the boundaries). The succession
of updating leading an unstable configuration
to a stable one is called an {\it avalanche}. Because of the dissipation at the boundaries,
all avalanches are {\it finite}. The structure of an avalanche can be encoded by the sequence of overcritical sites
$A = \left\{A_i\right\}_{0 \leq i}$ where $A_0 = \left\{a \right\}$, the activated site,
and  $A_i = \left\{j \in \Lambda | X_j \geq E_c \ \mbox{in the ith step
of avalanche}\right\}, i >0$.

The addition of energy is {\it adiabatic}. When an avalanche occurs,
one waits until it stops before adding a new energy quantum.
Further activations eventually generate a new avalanche,
but, because of the adiabatic rule, each new avalanche starts from {\it only one}
overcritical site. \\

Since the avalanche after activation of site $a$ maps overcritical to  stable configurations
one can view this process as a mapping
from $\calM \to \calM$ where one includes the process of activation of site $a$.
We hence associate a map $T_a$ with the activation at vertex $a$.
This map usually has singularities and therefore different domains of continuity denoted 
below by $M_a^k$ where $k$ runs through a finite set depending on $a$. Call 
$T_a^k = T_a \mid_{\calM_a^k}$. The main object of this paper is the study of the properties of 
the family of mappings $\{ T_a^k\}$ 
and to  link these properties to the asymptotic behaviour.

\ssu{Piecewise affine mappings.}
\sssu{Structure of the piecewise affine mappings.}

One can easily  write the 
 conditions on the stable energy configurations 
insuring that the avalanche $A = \left\{A_i\right\}_{0 \leq i}$ occurs.
This defines a convex domain\footnote{For example, the domain of energy such that a site a
do not relax after activation is delimited by the hyperplane $E_a = E_c-1$, and the boundaries of $\calM$.} $\calM_a^k$ in $\calM$
. The  $\calM_a^k$'s are the domains
of continuity of $T_a$ and they constitute, for each $a$, a partition of $\calM$. There is therefore
a one to one correspondence between an avalanche and a map $T_a^k$.  
The energy distributions rules of Zhang's model implies that:

\beq \label{Tij}
T^{k}_{a}.\bX = 
L^{k}_{a}.(\bX + {\be_a})= 
L^{k}_{a}.\bX + L^{k}_{a}.{\be_a}, \qquad \bX \in \calM^{k}_{a}$$ .
\eeq

\nid where the {\it linear} mapping $L^{k}_{a}$ characterizes the redistribution of
energies on each sites after the avalanche. The {\it column} $i$ of $L^{k}_a$'s 
contains the ratios of energy 
given by the site $j$ to the other sites after the corresponding avalanche. Alternatively, the entries
of the {\it row}
$i$ correspond to the energy received by $i$ from the others sites ($i$ included). $\be_a$ being the
canonical basis vector
of $\bbbr^N$
in the direction corresponding to the activation at site $a$, the constant
 vector $L^{k}_{a}.{\be_a}$ corresponds to the redistribution of the additional 
energy $\delta X = 1$ on each site, after the avalanche.
In the case where no relaxation occurs the corresponding map is just a shift along the $a$ axis.
A way to build $L_{a}^k$ is to construct it step by
step, by a left product of elementary matrices giving the redistribution of energy from
one step in the avalanche to the successive step.
 The composition of these matrices is determined by the avalanche profile.

Let $\calS_a^k = \partial{\calM_a^k}$. Then $\calS_a = \bigcup_k \calS_a^k$
is the the set of singularities for the transformation $T_a$. 
 The sets $\calS_a^k$ are unions of segments of hyperplanes
in $\bbbr^N$.

\sssu{Projection effect.} \label{Kernel}

The original Zhang's model contains a pathology.
Due to the reset to zero of an overcritical site after relaxation ($\epsilon =0$),
linear dependences among the sites (more precisely, direction of $\bbbr^N$ associated to these sites) 
are created along the avalanche. This implies the existence of a non trivial kernel.
Thus each $L_a^k$'s is a {\it projection}
onto a subspace of $\bbbr^N$,
whose dimension increases with the number of involved sites in the
avalanches.
The one step matrices have a number of zero eigenvalues
given by the number of sites set to zero at the corresponding time
step. Multiplying these matrices gives raise to the kernel of $L_a^k$.
We denote the subspace generated by zero eigenvalues $\calE^0(a,k)$.
Note that, in general, $KerL_a^k \subset \calE^o(a,k)$ .
Clearly, the existence of a nontrivial kernel is the source of several mathematical complications when studying the dynamics
of Zhang's model. It is a very particular feature of the $\epsilon=0$ model. 
 It makes however the global geometry of the attractor quite interesting (see Fig. \ref{F12}).

\sssu{Contraction}

Each mapping
$L_a^k$ 
has only eigenvalues of modulus lower or equal than $1$. 
Indeed, by definition, the sites of the boundaries of the avalanche receive
energy without relaxation.  This implies that, by eventually permuting the
basis vectors,  the linear map $L$ can  be written as:

\bea \label{splitting}
L= \left [\begin {array}{ccccccccccccc}
I &  *  \dots  * &  0\\
0 &  \left [\begin {array}{cccccc}
&  &   &  & \\
  &  &   &  &\\
\end {array}\right ] & 0\\
0  & *  \dots  * &  I\\
\end {array}\right ]
\eea

\nid where $I$ is the identity matrix and where the $*$'s can be zero or not. 
They correspond to the fraction of energy 
received by the sites which have not relaxed. Therefore, the  vectors
corresponding to sites not relaxing  are eigenvectors of $L$ with  eigenvalue one.
We denote the corresponding
(neutral) subspace by $\calE^n(a,k)$.

On the other hand, the inner block in (2) corresponds to the sites
which have relaxed.  The energy conservation implies that the sum on each 
column of the block is strictly lower than one (some part of the energy has gone outside the block, 
to the sites
on the boundary of the avalanche). By usual arguments on positive matrices it follows that 
the eigenvalues are {\it strictly lower than one} in the block \cite{Gantmacher}.
Note that, for $\epsilon =0$, this block contains also the subspace 
$\calE^0(a,k)$. Therefore, the subspace of relaxing sites is decomposed 
into two subspaces: $\calE^o(a,k)$ and 
$\calE^-(a,k)$, where $\calE^-(a,k)$ denotes the subspace associated to the eigenvalues $0<|\lambda_i| < 1$.

Hence to each mapping $L_a^k$ we associate the following decomposition:

\beq \label{subs_split}
\calE^-(a,k) \oplus \calE^n(a,k) \oplus \calE^0(a,k)
= \bbbr^N. 
\eeq

\ssu{Composed mapping.} \label{Prod_map}

\sssu{Composition of affine mappings. Extended dynamical system.}

The activation dynamics can be 
represented by the {\it left Bernoulli
shift}  $\sigma$ over $\Sigma^+_\Lambda$, the set
of right infinite sequence $\ba = \left\{ a_1,\dots, a_k, \dots\right\},
a_k \in \Lambda$, where $\sigma \ba = a_2a_3 \dots $. Namely, $a_n$ is the $n$ th activated
site in the activation sequence $\ba$.
We denote by  $\left[ a \right]$  the set of sequences whose first digit is $a$.

The combined effect of the activation and relaxation process 
is than described by a dynamical system of skew-product type
$\calT : \Omega \longrightarrow \Omega$ such that:
$$ \calT(\hat{\bX}) \deq \left( \sigma \ba, T_{a_1}(\bX) \right) \ ; \qquad 
\hat{\bX} 
{\stackrel {\rm def}{=}} (a,\bx)
$$
where  $\Omega =
\Sigma^+_\Lambda \times \calM$ is called the extended phase space.

Let $D \calT_{\hat{\bX}}$ be the tangent map of $\calT$ at $\hat{\bX}$. (When speaking about differentials of $\calT$ we usually think of $\Sigma_\Lambda$ represented by a smooth system $z \to |\Lambda|\cdot z \mbox{mod}\ 1$).

The singularity set of $\calT$ is :

\beq \label{S} 
\calS = \bigcup_{a \in \Lambda} \left[ a \right] \times \calS_a
\eeq

We define a distance on $\Omega$ by $\D{d_{\Omega}(\hat{\bX},\hat{\bY}) = 
d_{\Sigma^+_\Lambda}(\ba,\ba') + d_{\calM}(\bX,\bY)}$, where $\hat{\bX} = \left(\ba, \bX \right), \hat{\bY} = 
\left(\ba', \bY \right)$.
We denote the two projections on the first and second coordinate by $\pi^u \left(\ba, \bX \right) = \ba$, and $\pi^s
\left(\ba, \bX \right) = \bX$. The superscript $u,s$ means respectively {\it unstable}
and {\it stable} and will be explained below. 
We have a natural partition of $\Omega$, $\calP = \left\{ \calP^k_a =
\left[ a \right]  \times \calM^k_a\right\}$. 
Note that $\calP$ is a generating partition for $(\calT,\Omega)$
in the topological sense, that is, the diameter of 
the elements of $\bigvee_i \calT^{-i} \calP$ goes to zero. \\

\sssu{Kernel of the infinite product map.}

For $\epsilon=0$ the kernel of the map $\calT^t$ can increase with $t$, projecting $\bbbr^N$
onto spaces of lower and lower dimensions.
Therefore, after a certain,
{\it finite time}, $n(\hat{\bX})$,
$\hat{\bX}$ is projected onto the effective lower dimensional
subspace \footnote{One has to check that this space is not finally reduced 
to $\left\{0\right\}$ ! However, this would imply that all vectors in $\bbbr^N$ are asymptotically mapped
to 0. This is excluded since the avalanche matrices have non negative entries.}
\footnote{Note that the dimension of this space depends {\it a priori} on $\hat{\bX}$.
However, if ergodicity holds, this dimension is constant for (almost-every) initial condition.}: 

\beq \label{E_eff}
\calE^s(\hat{\bX}) \deq 
 \left\{\bv \in \left\{ 0 \right\} \times  \bbbr^N	 \ ; \ \forall t \geq 0, 
\|D\calT^t_{\hat{\bX}}.\bv \| > 0 \right\}
\eeq

It is somehow the reference space with respect to $\hat{\bX}$, because,
asymptotically, the dynamics of vectors in $\bbbr^N$ under
$D\calT_{\hat{\bX}}$  reduces to the dynamics of vectors initially in
$\calE^s(\hat{\bX})$. We get therefore a splitting of the projection on $\calM$
of the tangent space at $\hat{\bX}$ as:

\beq
\bbbr^N = \calE^s(\hat{\bX}) \oplus \calK(\hat{\bX})
\eeq

\nid where $\calK(\hat{\bX})$  is the kernel of the product map. This
splitting will be  refined further below by using the Oseledec space
decomposition.

\sssu{Local contraction.}

One easy shows that for any finite connected $\Lambda$ and arbitrary activation sequence $\ba$ all sites become overcritical infinitely often.
Assume the opposite. Then, there exists a site which is overcritical only a finite number of times
along the infinite sequence $\ba$ but has a neighbour site which relaxes infinitely often. The energy
coming from the overcritical neighbour site is larger than
$\frac{(1-\epsilon)E_c}{2d}$ by definition. This implies that all neighbours relax also
an infinite number of times during the whole sequence.
Hence we get a contradiction. It follows that there exists a time 
$\tau \equiv \tau(\Lambda,E_c,\epsilon) < \infty$
such that, $\forall \hat{\bX}$,  after at most $\tau$ 
time steps each site has been at least once overcritical. 
By looking at the product  map $D\calT^\tau_{\hat{\bX}}$ this implies that all eigenvalues
are different from one. This is straightforward since the sum of entries on each column
of the composed map on $\calM$
is strictly lower than one (and is  bounded away from 1).
Therefore there is a positive constant $C \equiv C(\Lambda,E_c,\epsilon)$ s.t.:

\beq \label{Contract}
\| \pi^s D\calT^\tau_{\hat{\bX}} \|_1 =
\sup_{\| V \|_1 =1} \| \pi^s D\calT^\tau_{\hat{\bY}} V  \|_1 < C <1\ .
\eeq

This implies that the map $\calT^\tau$ acts as local contraction  in all directions 
in the space $\calM$, along the trajectory
of any point $\hat{\bX}$ .
 This has in particular the following consequence. The distance of two points
$\hat{\bX},\hat{\bY}$ whose trajectory belong to the same domain of continuity
eventually goes to zero if the trajectories
lie in the same domains of continuity along the {\it whole} activation sequence.

\sssu{Hyperbolic structure and Lyapunov exponents.}

Assume that almost every point is regular, namely the map $\calT$
is differentiable along all points of the trajectory (note that as long we work in the tangent spaces this assumption is not necessary since the involved mapping are all well defined at $\partial P_i$. Only for the construction of the local induced stable manifolds in $M$ one has to take care of regularity). Then, one can decompose the (effective)
tangent
space at a.e. point  $\hat{\bX} \in \Omega$
into a contracting subspace $\calE^s(\hat{\bX})$ and an expanding
one $\calE^u(\hat{\bX})$, s.t. :

\ben
\item $\forall \hat{\bX}, \ \calE^s(\hat{\bX}) \oplus \calE^u(\hat{\bX}) = \bbbr^{N+1-dim(\calK(\hat{\bX}))}$.

\item $\forall \hat{\bX}, \ \exists \lambda < 1, \tau < \infty$, 
s.t. $\| D \calT^\tau \|_{\calE^s(\hat{\bX})} \leq \lambda$. Furthermore $\| D
\calT \|_{\calE^u(\hat{\bX})} = N = |\Lambda|$.

\item $\calT(\calE^s(\hat{\bX})) = \calE^s(\calT(\hat{\bX})) ; \ \calT(\calE^u(\hat{\bX})) = \calE^u(\calT(\hat{\bX}))$
\een

Furthermore $\calE^s(\hat{\bX})$
can be decomposed into a sequence of subspaces \cite{ER,Pollicott,Young} \footnote{In fact,
in the $\epsilon = 0$ case we have still to assume that the angle between
$ \calE^s(\hat{\bX})$ and $\calK(\hat{\bX})$ is bounded away from zero, a.s. because,
otherwise, Lyapunox exponents might not exist}:

\beq
\calE^s(\hat{\bX}) = \calE_{1}(\hat{\bX}) \supset \calE_{2}(\hat{\bX}) \supset \dots \supset
\calE_{l}(\hat{\bX})
\eeq

\nid such that if $\bv \in \calE_{i}(\hat{\bX}) \setminus \calE_{i+1}(\hat{\bX})$
the average contraction of $\bv$ is given
by the Lyapunov exponent :

\beq \label{Lyapunov}
\lambda_i(\hat{\bX}) =
\lim_{n \rightarrow \infty} \frac{1}{n} log \|D\calT^n_{\hat{\bX}}(\bv) \|_2
\eeq

\nid From property
(\ref{Contract}) there are no zero Lyapunov exponents (note
however that  some exponents go to zero as $E_c$ tends to infinity). 

If the dynamics is ergodic
the  Lyapunov exponents are almost-surely constants and the same holds for $ \dim \calE_i$. 
Corresponding to the shift action
there is  a {\it positive} Lypunov exponent, which is trivially
$log(N)$.
The Lyapunov exponents are directly related to the geometrical structure of the support of 
the invariant measure.
In the Zhang's model the negative exponents $0 > \lambda_1 \geq \lambda_2 \geq \dots \geq 
\lambda_N$
are physically related to the transport of energy and the dissipation rate
at the boundary \cite{BCK3}. In particular, we show below that there is a natural relation 
linking these
exponents to the avalanche size distribution. The positive Lyapunov exponent $\lambda_0 = logN$
corresponds to the entropy production coming from the activation dynamics.

 The average exponential volume contraction rate on $\calE^s(\hat{\bX})$ is given by the sum:

\beq \label{Contrac_vol}
\sum_{i=1}^N\lambda_i
\eeq

\nid while the average exponential variation  rate of the volume in the extended phase space is
$logN + \sum_{i=1}^N\lambda_i$.\\

For regular $\hat{X}$ let 
$$
\calW^s_\epsilon (\hat{\bX}) =
 \left\{ \hat{\bY} : d (\hat{\bX}, \hat{\bY}) \le \epsilon \
\forall i \ge 0,\ \ \calT^i \hat{\bX}\ ,\right.
$$
$$\left. \mbox{and}\ \calT^i \hat{\bY}\ \mbox{are\ in\ the\ same\ partition\ element\
of}\  \calP \right\}\
$$

be the $\ \epsilon$ local stable manifold.
Clearly one has on $\calW^s_\epsilon$ uniform exponential contraction.

The global stable manifold $\calW^s (\hat{\bX})$ is obtained by
\[
\bigcup_{i\ge 0} \calT^{-i} (\calW^s_\epsilon (\hat{\bX})\ .
\]
Finally let $\calW_{loc} (\hat{\bX})$ be the largest connected component of $\calW^s$ containing $\hat{\bX}$.
Since the system is of skew product type one has a trivial unstable manifold being in the case of
 representing the shift as $z \to z \cdot |\Lambda | \mbox{mod}1$ the whole interval $\left[ 0,1 \right]$.
Note that 
$\calW_{\epsilon}^s(\hat{\bX})$ may not exist 
if $\exists
\left\{n_i \right\}$ s.t. $d(\calT^{n_i}(\hat{\bX}),\calS) < e^{-{n_i}C}$ 
where $C>0$ is some constant 
larger
than $-\lambda_1$. The set of points with this property has measure zero
unless the invariant measure concentrates on $\calS$. 
This aspect will be described in more detail in \cite{BCK2}. We make the following
conjecture:

 \bconj \label{IFS1}
There exists a $\bar{E_c}(N)$, such that, for Lebesgue almost-every
 $ E_c < \bar{E}_c(N)$
there exists an $n(E_c,N)$ and a $\nu$ such that
$\forall t > n(E_c,N)$: %
$$d(\calT^t(\Omega),\calS) > \nu > 0$$
\econj

This implies that after a finite time the dynamics  stays away from the singularity set.
 This assumption is sufficient for the existence of local stable manifolds, but it will
have several other important implications. We expect Conj. \ref{IFS1}  to
be true for $\overline{E}_c$ sufficently small since for $\overline{E}_c \ll 1$ the contraction dominates the
expansion in the extended phase space. In this case the invariant set has
the structure of a totally disconnected Cantorset with large gaps. Furthermore the local structure
of this invariant set is constant for open sets of $E_c$ values
(see section 6), but the singularity set is varying continuously in $E_c$ (except for a countable set),
one can find open domains of $E_c$ values where $\calS$ stays away from the invariant set.\\

The singularity set has nevertheless the following effect on the dynamics.
Take an $\eta$-ball of initial conditions (in $\calM$),
and fix an activation sequence.
For $\eta$ large enough the image of the ball under some iterate
of $\calT$ is cut by the singularity set. This
means that the points separated by the singularity set will not evolve under the same sequence of mappings. For large size systems this can cause on $\calM$ a kind of expansion effect (for fixed typical activation sequence) on a mesoscopic scale.

\sssu{Symbolic dynamics.}

Symbolic dynamics is a very useful tool for the investigation of the orbit
structure of dynamical systems. To this aim, one fixes a partition $\calP$ of the phase
 space and associates to each point the sequence of partition elements visited by
the orbit of a point. To make symbolic dynamics useful one wants this
correspondence essentially to be unique (that is up to sets of measure zero).
Furthermore to handle the symbolic dynamics it is essential to have an
explicit characterization of the legal (that is by orbits generated) set of
symbolic sequences. The perhaps most prominent example of such an explicit
description are symbolic systems defined by a Markov transition graph called
Subshift of Finite Type (SFT). A classical result in hyperbolic dynamics says
 that uniform hyperbolic systems are always conjugated to SFT \cite{Bowen,Ruelle,Sinai}.
The specific partitions giving rise to such coding are called Markov Partitions.

In the Zhang's model one can  encode the possible transitions between avalanches
in a {\it transition graph} with respect to the canonical partition
$\calP = \left\{ P^k_a \right\}$.
Namely, we draw an arrow from $\calP_a^k$ to $\calP_b^l$ if and only if 
$\calT(\calP_a^k) \cap \calP_b^l \neq \emptyset$.
We denote by $\Sigma^+_{\calP}$
the set of admissible infinite sequences w.r. to the partition $\calP$. .

Clearly, points on $W_{loc}^s(\hat{\bX})$
form an equivalence class for the symbolic coding induced by $\calP$. Note that the transition graph is a priori Markov only for special choices of $E_c$.  


%

When the invariant set is bounded away from the singularity set one can refine the partition 
$\left\{\calP_a^k \right\}$ to make it Markov (this is certainly not a necessary
assumption to get a Markov transition graph).
Namely, there is an $m$ s.t.
the partition $\bigvee_{i=1}^m \calT^{-i}(\calT^m\calP)$ is a Markov partition.
We label the affine mappings corresponding to the Markov partition elements by $F_i$. Note 
that several $F_i$ can usually correspond to the same map $T_a^k$.

Let us give an example. In the case $E_c \in [1,2]$, $\epsilon =0$, in one dimension,  the piecewise
continuous mapping applied
is {\it uniquely} determined by the position of the zero site.  Indeed, after a sufficiently long sequence
all sites have energy $X_i \geq
\frac{E_c}{2}$
but eventually one with a zero value. 
Since a site with value zero is the only possible stopping site for an avalanche besides the 
boundary the avalanche is {\it uniquely determined}
by the position of the activated site and of the zero.
This case is however the simplest, because there is no need to cut further the
$\calP_a^k$'s in order to get a SFT. For $E_c > 2$ things are more complicated, due to the 
presence
 of sites
with integer values $1 \dots [E_c-1]$ which may stop an avalanche, according to the amount 
of energy they receive.

As already said we expect the above mentioned property of being disjoint from the
 singularities to hold for a.e. $E_c$ value less than some $\bar{E}_c(N)$.
This implies that the system is a SFT. For $\epsilon=0, d=1 $ there is another dense
 set of $E_c$ values
for which one can show that the system is a SFT.

\bp
For $\epsilon=0$, $d=1$ and $E_c=n/(2d)^p$ for any $n,p \in \bbbn^*$ the system is conjugate to
a SFT
\ep   

\medskip
\noindent
Note that the elementary operations on each avalanche and each node $i$ are of the form
$X_i \rightarrow X_i + \sum_{j}\frac{X_j}{2}$ for some  $j$ and a check wether $X_i$
is larger or less than $E_c$. If $E_c$ is of the above form it is a finite digit number
 (eventually
zero) in base $2$. It follows that for each point $\bX \in \calM$ one
 has to know only a uniformely
bounded, finite number of digits in base $2$ to decide in which set $\calM_a^k\ $ $\ \bX$ is. 
The same holds for the legal transitions between avalanches domains, that is, there is a finite 
number
of forbidden strings in base $2^{|\Lambda|}$ coding the whole system, hence it is a SFT.
\epr

\sssu{Macroscopic state and SBR measures.} \label{Measure}

The addition of energy on one hand, and the dissipation
of exceeding energy at the boundaries, on the other hand, drives gradually the system towards a stationary state where there is a constant
energy flux through
the system.  As far as our representation accounts for activation dynamics on one hand and
transport-dissipation (avalanche) on the other hand,
the full informations about the macroscopic behaviour of the system at stationarity is contained in
the {\it invariant measures}\footnote{Namely $\mu(\calT^{-1}(\calB) = \mu(\calB)$ where $\calB$ is a 
measurable set in $\Omega$.} of our dynamical system. 
Since $\Omega$ has a product structure one has canonical measures $\mu^u$ (induced measure on
the unstable direction) and $\mu^s$ (induced measure on $\calM$). For simplicity we will assume
that $\mu$ is a Bernoulli measure, namely that the sites are chosen independently
with fixed rates.
Once we have fixed the distribution
 of activation, we  are interested on the possible $\mu^s$ measures.
Of special physical importance are the measures obtained by iterating the Lebesgue measure $\mu_L$ 
on $\calM$, that is $\lim_{n \rightarrow \infty} \frac{1}{n}\sum_{i=0}^{n-1} \calT^{i}(\mu^u \times \mu_L)$. We call this measure conditional SRB with respect to $\mu^u$.

It is common in the SOC litterature to assume ergodicity. In our setting the physically relevant 
ergodic property is equivalent to the following conjecture.

\bconj \label{irreduc}
For any $E_c,\Lambda,\epsilon$, and given $\mu^u$ the corresponding conditional SBR measure is unique.
\econj 

This implies in particular the almost-sure equality between the ensemble average and the time average
for typical energy configurations.
We give some arguments to support this assumption at least for certain
$E_c$ values.
If the probability of activation of any site in non zero then there exists a periodic point
 $\hat{\bX}$ with period $p$ such that
\beq \label{mucyl}
\mu^u\left(\left[\pi^u(\hat{\bX}),\pi^u(\calT(\hat{\bX})), \dots,\pi^u(\calT^{p-1}(\hat{\bX}))\right]\right)
> 0
\eeq

\nid where $\left[\pi^u(\hat{\bX}),\pi^u(\calT(\hat{\bX})), \dots,\pi^u(\calT^{p-1}(\hat{\bX}))\right]$ 
is a cylinder set.
This is the set of infinite sequences in $\Sigma_\Lambda^+$
which coincide with the activation sequence of $\hat{\bX}$
on the $p$ first symbols. Assume  that the periodic orbit admits a stable manifold such that

\beq \label{toptrans}
\bigcup_i \calW^s(\calT^i(\hat{\bX})) = \calM
\eeq
If the periodic point does not lie on $\calS$ one can take a small neighbourhood $\calU_\epsilon(\hat{\bX})$.
Otherwise, as far as the singularity set is moving with $E_c$ while the limit cycle
does not change on open domains of $E_c$ (the maps remain the same) one can change $E_c$ by an 
arbitrary small
value in order to make the limit cycle disjoint from $\calS$. Due to (\ref{mucyl}),
a generic sequence $\ba$ admits
arbitrary long segments with repeated words $\pi^u(\hat{\bX}),\pi^u(\calT(\hat{\bX})), \dots,\pi^u(\calT^{p-1}(\hat{\bX})$. Therefore, from (\ref{toptrans}), 
almost every points visits $\calU_\epsilon(\hat{\bX})$  infinitely often 
for any $\epsilon>0$ sufficiently small. By the hyperbolic structure and some moderate assumptions on the 
distribution of the size of $\calW^s_{loc}(\hat{\bX})$ one can then form a Hopf chain\footnote{A path
made of pieces of local stable and unstable manifolds.} between iterates of a.e.
points $\hat{\bZ},\hat{\bY}$ when they visit $\calU_\epsilon(\hat{\bX})$. By standard arguments from ergodic theory
concerning the equality of forward and backward averages one can then prove that a.e. pair of points on the 
invariant set belongs to the same ergodic component. In general, it does not seem  easy to give explicit 
examples of sequences of avalanches satisfying the above conditions. The crucial point here is to show
(12). But perhaps it should be possible to weaken the above conditions substantially.

For $d=1, E_c > 1$, one can check it by using the following argument.
 For $E_c > 1$, starting from any stable configuration,
one can add energy to the low energy sites ($E_i < E_c -1$)
in order to get a configuration where all sites have energy
$E_c -1 < E_i < E_c$. Activating any site in this configuration generates
a unique "maximal" avalanche where all sites become overcritical. This avalanche is  recurrent
\footnote{This argument has been already used by other authors
like Dhar \cite{Dhar}, and Speer \cite{Speer} for the Dhar model.}
and there exists a periodic orbit satisfying (12).
For $d >1$ the number of reflexions
of the front on the boundaries can vary and there are several types of "maximal" avalanches
which makes the argument break. One can however still apply it on a diamond shaped lattice with 
with $L$ odd ($N=L^2$) , because, by activating periodically in the middle site,
one has essentially the same situation as for the one dimensional chain.  
For $E_c$ small,  especially
for $E_c <1$ the avalanche patterns are much more complicated and the above argument breaks down.

Note that in the case where we have a Markov graph,
 the ergodic property can in principle be directly checked on the Markov transition 
graph defined in the previous section. Namely, if the Markov transition graph is asymptoticaly irreducible
 and aperiodic,
(unique) ergodicity follows from usual results on Markov
chains. \\

We proceed in discussing some aspects of the dependence of $\mu$ on $E_c$ for fixed $\Lambda$.

\bp \label{musing}
$\mu^s$ is singular for all $E_c$ sufficiently small.
\ep

\bpr
This follows easily since for $E_c <<1$ one can make the $L_1$ norm of all avalanche map arbitrary
small since the avalanche has to ``reflect" many times on each boundary node, hence every node has contributed to the dissipation. Since the expansion is constant it follows that $det(D\calT) < 1$ hence 
all measures are singular.  
\epr

\bp
$\mu^s$ is atomic for the chain and $E_c \in
[\frac{1+\epsilon}{1-\epsilon},\frac{2}{1-\epsilon}] $.
\ep

This is proved in section III.A.3.
Furthermore, we conjecture the following:

\bconj
The Hausdorff dimension of $\mu^s$ is piecewise continuous and monotonously increasing on the domains
of continuity for $E_c <<1$.
\econj

This is supported by the following argument. On open intervals $I_i$ of $E_c$ the structure of the mappings 
$T_a^k$ does not change but the domains of continuity $\calM_a^k$ do. Furthermore for $E_c$ decreasing the 
probabilities for avalanches with higher contraction should  increase which should force the Hausdorff
dimension to increase monotonously with $E_c$ on each $I_i$.\\

We now discuss the connection between the invariant measure and the SOC state.
It is possible to extract from $\mu$ the probability
distribution of all observables usually considered in the study of SOC.
The traditionaly used
observables are: the  {\it duration} $t$ (number of iteration steps inside one avalanche);
the {\it size} $s$ ({\it total} number of relaxing sites counted with multiplicity), and the {\it area}  $a$
(number of {\it distinct} relaxing sites).
Fix now an observable, say $s$. Let $\calK_s$ be the set of mappings $T^k_a$ with avalanche size $s$ and let $\calQ_s$ be the union of it's domains $\calM^k_a$.
Let  $P_N(s)$  be the probability
to have an avalanche of size $s$ for a lattice of size $N$ in the staionary limit. One has
clearly:
\beq
P_N(s) = \mu^s(\calQ_s)
\eeq

\sssu{Ledrappier-Young Formula.}  \label{LY_sec}
This formula plays a key role in relating the probability of
avalanche size to the average contraction rate (sum of Lyapunov exponents).
 It establishes a kind of conservation
law relating the Lyapunov exponents, some version of Haussdorf dimension and the Kolmogorov-Sinai
entropy.

One  can refine the foliation  into
a stable and unstable manifold by splitting the 
manifolds
into sub-manifolds $\calW^s_i(\hat{\bX})$ (resp $\calW^u_i(\hat{\bX})$)
such that the contraction (resp. the expansion) on $\calW^s_i(\hat{\bX})$ (resp.
$\calW^u_i(\hat{\bX})$ ) is governed by the Lyapunov exponent $\lambda_i$.
 Let $\delta_i$ be the local Haussdorf dimension of the measure
$\mu$ projected on $\calW^r_i(\hat{\bX})$ (where $r$ stands for $s,u$), namely:

\beq \label{HDi}
\delta_i = \D{lim_{\epsilon \rightarrow 0} \frac{log \mu(\calB_i(\hat{\bX},\epsilon))}{log\epsilon}}
\eeq

\nid where $\calB_i(\hat{\bX},\epsilon))$  is an $\epsilon$-ball
around $\hat{\bX}$ in  $\calW^r_i(\hat{\bX})$. Then $\sigma_i = \delta_{i} - \delta_{i+1}, i=1 \dots N-1$
  is the
transverse dimension of the measure $\mu$ on $\calW^r_{i}(\hat{\bX}) \setminus \calW^r_{i+1}(\hat{\bX})$. It is  constant
for $\mu$ almost-every  $\hat{\bX}$  if  $\mu$ is ergodic.
The unstable foliation being one dimensional in our context, the
Haussdorf dimension of the measure $\calW^u(\hat{\bX})$ is $\delta_0$.
It is equal to one for  the uniform activation measure.

Let $h_{\mu}$ be the Kolmogorov-Sinai entropy of $\mu$ and
$\lambda_i^+$ the positive Lyapunov exponents. The Ledrappier-Young formula is \cite{LY} (for ergodic measures):

\beq \label{LY_forw}
h_{\mu}= \sum_{i} \lambda_i^+ \sigma_i
\eeq

\nid where the sum is taken over the positive Lyapunov exponents. It
expresses in particular that
without any absolute continuity of $\mu$, any equation relating entropy and positive Lyapunov exponents must involve some notion of fractional dimension.
In our case, it reduces to :

\beq
h_{\mu}= logN \delta_0
\eeq

From now on we will assume that  $\mu^u=\mu_L$ (uniform activation). In this case $\delta_0=1$.

When the dynamics is invertible,
this formula, applied to the inverted system, gives the following equality in the Zhang's
model:

\beq \label{LY}
\sum_{i=1}^N\ \lambda_i\ \sigma_i = -logN
\eeq
where the sum is now taken over the negative Lyapunov exponents.

However, one has to assume that the dynamics is ($\mu$ almost-surely)
 invertible.
 That  means physically  that, at stationarity,
 the probability that two avalanches, starting
from two different
configurations, end on the same configuration of energies is zero.
Like conjecture \ref{IFS1} we expect this property to hold only for small $E_c$ values,
where the invariant set
is a Cantor set  but to fail
generically for large $E_c$ values.

We have the following conjecture:

\bconj \label{Invertibility}
For $E_c$ sufficiently small, there exists a  $n(E_c,N,d)$
 such that $\forall t> n(E_c,N)$:
$$\mu(\calT^t(\calP^k_i) \cap \calT^t(\calP^l_j)) = 0, \ \forall P^k_i \neq P^l_j $$

\econj

Note that one can  weaken this assumption
by requiring that there are, on the attractor, less than $N$ preimages and still get a nontrivial
relation to Lyapunov exponents. One
can still write down a Ledrappier-Young formula for non invertible systems by making the system invertible \cite{Schmeling}
by coding the backward iteration tree in the same way as we did with the activation sequences,
hence introducing an additional variable on which the forward dynamics contracts. Let  $J_N(\hat{\bX})$ be the
number of preimages of $\hat{\bX}$ and $J _N= \int J_N(\hat{\bX}) d\mu(\hat{\bX})$
the averaged number then:

\beq \label{LYgen}
- \sum_{i=1}^N \lambda_i \sigma_i = logN - logJ_N
\eeq

\su{Dynamics and SOC.}

\ssu{The Zhang's model as an iterated function system.}

If the system is conjugate to a subshift of finite type, the dynamics of the Zhang's model
is essentially equivalent to a graph probabilistic Iterated Function System (IFS) 
\cite{Barnsley,Falconer},
namely, a set of quasi-contractions $F_i$ randomly composed along a Markov graph admitting
a unique invariant measure $\mu^\ast$. Note that IFS are usually defined for true contractions, however, in our case,
any finite composition along the graph is a contraction.
In this case, the classical theory of graph directed
Iterated Functions Systems  applies and allows one to obtain interesting results
 with respect to the geometrical structure
of the invariant set.

\sssu{The Zhang's model attractor.}

The IFS determines a unique non-empty
compact set $\calA$, called {\it the attractor of the IFS},
satisfying:

\beq
\calA=\calF(\calA) \deq\D{\bigcup_{i=1}^{\calR_N} F_i(\calA)}
\eeq

This set is usually a fractal.

Let $\bbbh(\calM)$ be the set of compact subsets in $\calM$.
Define a distance on $\bbbh(\calM)$, called the {\it Haussdorf metric}  by:

\beq
\delta(A,B) = sup \left\{ d(a,B), d(b,A), a \in A, b \in B \right\}
\eeq

\nid where $A, B$ are non empty closed bounded subsets
of $\calM$, and $d(x,A) = inf \left\{ d(x,a), a \in A\right\}$.
$\calA$ is an attractor of the IFS in sense that it satisfies the following property
\cite{Hutch} :
$$\forall \calB \in \bbbh(\calM), \calF^n(\calB) \rightarrow \calA$$
\nid in the Haussdorff metric
when $n \rightarrow \infty$. Furthermore, if  $\calB \in \bbbh(\calM)$ is such, that for all $i$, $F_i(\calB) \subset \calB$
then :
$$\calA= \bigcap_{n=0}^{\infty} \calF^n(\calB)$$

Therefore, the asymptotics dynamics of the Zhang's model lives onto an attractor, further on denoted by $\calA$,
whose fractal geometry  is linked to the critical behaviour at stationarity.
Note however, that, despite one might expect from the presence of dissipation  the
existence of an attractor with  a fractal structure in general SOC models,
this is not the case because the distribution  of energy has to be of type
like in the Zhang's model to get local contraction effects.

We give now two simple examples of attractors which can be constructed "by hand" .

\sssu{One dimensional chain with $E_c=1, \ N=3, \epsilon = 0$.}

For $N=3$, each configuration $\bX$ is a triplet $\left\{ X_1, X_2, X_3 \right\}$.
First note that only the mappings whose image
intersect the cube $[\frac{E_c}{2},E_c[^3$ are relevant for the asymptotic dynamics.
Moreover, for $E_c \leq 1$ each activation generates an avalanche,  and
the
resulting configuration always contains a site with zero energy. This is an effect
of projection onto the complementary set of the kernel of the product mapping, discussed
in section \ref{Prod_map}.

The mappings   (rather, their projection onto the faces of $\calM$)
are respectively:

\bea \label{map3}
F_1 = \left[ \begin {array}{ccc}
\frac{1}{2} & 0 \\
\frac{1}{4} & \frac{1}{2} 
\end {array}
\right]
.  \left[\begin {array}{cc} X \\
Y
\end {array}\right]
+ \left[\begin {array}{cc} \frac{1}{4} \\
 \frac{1}{8}
\end {array}\right], \
\nnb F_2  = \left [\begin {array}{ccc}
\frac{1}{2} & 0 \\
\frac{1}{4} & \frac{1}{2} \\
\end {array}
\right ]
.  \left [\begin {array}{cc} X \\
Y\\
\end {array}\right ]
+ \left [\begin {array}{cc} \frac{1}{2} \\
 \frac{1}{4}\\
\end {array}\right ], \
\nnb F_3 = \left [\begin {array}{ccc}
\frac{1}{2} &  \frac{1}{4}\\
\frac{1}{2} & \frac{1}{4} \\
\end {array}
\right ]
.  \left [\begin {array}{cc} X \\
Y\\
\end {array}\right ]
+ \left [\begin {array}{cc} \frac{1}{4} \\
 \frac{1}{4}\\
\end {array}\right ]\\
\nnb F_4 = \left [\begin {array}{ccc}
\frac{1}{2} &  0\\
0 & 1 \\
\end {array}
\right ]
.  \left [\begin {array}{cc} X \\
Y\\
\end {array}\right ]
+ \left [\begin {array}{cc} \frac{1}{2} \\
 1\\
\end {array}\right ],\
\nnb F_5 = \left [\begin {array}{ccc}
\frac{1}{4} & \frac{1}{4} \\
 \frac{1}{4}& \frac{1}{4} \\
\end {array}
\right ]
.  \left [\begin {array}{cc} X \\
Y\\
\end {array}\right ]
+ \left [\begin {array}{cc} \frac{1}{4} \\
 \frac{1}{4}\\
\end {array}\right ], \
\nnb F_6 = \left [\begin {array}{ccc}
1 &  0\\
0 & \frac{1}{2} \\
\end {array}
\right ]
.  \left [\begin {array}{cc} X \\
Y\\
\end {array}\right ]
+ \left [\begin {array}{cc} \frac{1}{2} \\
 1\\
\end {array}\right ]\\
\nnb F_7 = \left [\begin {array}{ccc}
\frac{1}{4} &  \frac{1}{2}\\
\frac{1}{4} & \frac{1}{2} \\
\end {array}
\right ]
.  \left [\begin {array}{cc} X \\
Y\\
\end {array}\right ]
+ \left [\begin {array}{cc} \frac{1}{4} \\
 \frac{1}{4}\\
\end {array}\right ], \
\nnb F_8 = \left [\begin {array}{ccc}
\frac{1}{2} &  \frac{1}{4} \\
0 & \frac{1}{2} \\
\end {array}
\right ]
.  \left [\begin {array}{cc} X \\
Y\\
\end {array}\right ]
+ \left [\begin {array}{cc} \frac{1}{4} \\
 \frac{1}{2}\\
\end {array}\right ], \
\nnb F_9 = \left [\begin {array}{ccc}
\frac{1}{2} &  \frac{1}{4}\\
0 & \frac{1}{2} \\
\end {array}
\right ]
.  \left [\begin {array}{cc} X \\
Y\\
\end {array}\right ]
+ \left [\begin {array}{cc} \frac{1}{8} \\
 \frac{1}{4}\\
\end {array}\right ]
\eea

The Markov transition graph can be easily computed.
Each legal transition occurs with probability $\frac{1}{3}$ (activation of sites 1,2,3).
To obtain the invariant set of the IFS, one must first notice  that the three mappings $F_3,F_5,F_7$
have a zero eigenvalue and project vectors in $\bbbr^3$ along  the direction  $\left [\begin {array}{cc} 0 \\1 \\
-2\end {array}\right ], \ \left [\begin {array}{cc} 1 \\0\\
-1\end {array}\right ] , \ \left [\begin {array}{cc} -2 \\
1 \\ 0 \end {array}\right ]$. These projection induce a tree  structure for the invariant set.
The maps send their domain of continuity onto the segments:

$$a =  \left\{  \bX \in \bbbr^3 \  | \ \bX = \lambda. \left [\begin {array}{cc} 0 \\ 3/4 \\
1/2 \end {array}\right ]+ (1- \lambda). \left [\begin {array}{cc} 0 \\1 \\
1 \end {array}\right ], \lambda \in [0,1]  \right\}$$

$$\ b = \left\{ \bX \in \bbbr^3 \ | \ \bX = \lambda. \left [\begin {array}{cc} 1/2 \\ 0 \\
1/2 \end {array}\right ]+ (1- \lambda). \left [\begin {array}{cc} 1 \\0 \\
1\end {array}\right ], \lambda \in [0,1] \right\}$$

$$ c = \left\{\bX \in \bbbr^3 \ | \ \bX = \lambda. \left [\begin {array}{cc} 1/2 \\3/4 \\
0 \end {array}\right ]+ (1- \lambda). \left [\begin {array}{cc} 1 \\1 \\
0 \end {array}\right ], \lambda \in [0,1] \right\}$$

We can generate the invariant set
by starting with the set : $a \cup b \cup c$. We show in Fig.~3 the
initial branches $a, b, c$ and their image under
the
five first iterates of the IFS. We have labeled the branches of the tree
by their corresponding coding, for the three first iterates. One see, then how the tree structure is generated.\\

\bef
 \bc
 \begin{minipage}{8cm}
  \epsfxsize=8cm
 \epsfysize=6cm
 \epsffile{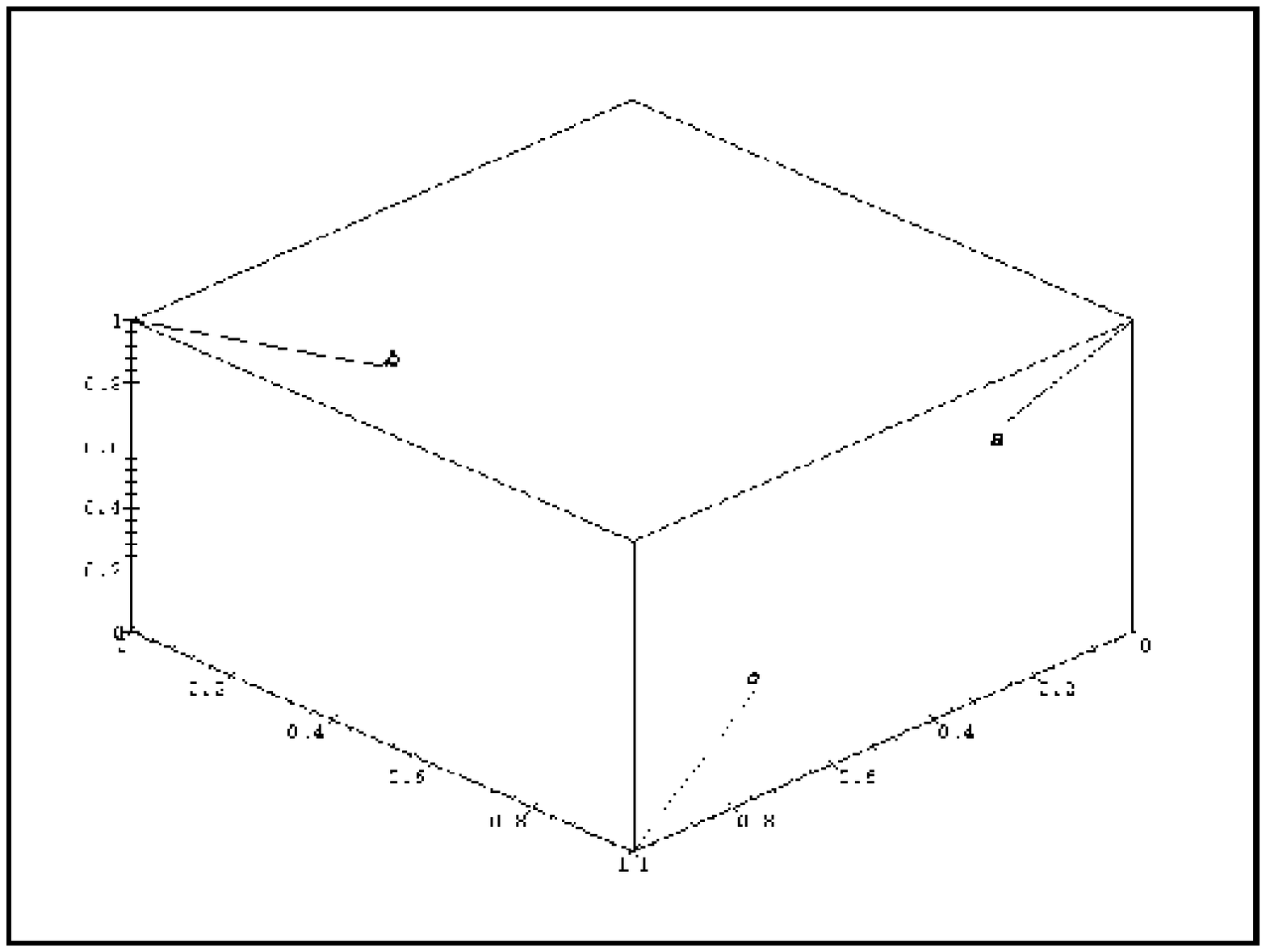}
\end{minipage}
\hspace{1cm}
 \begin{minipage}{8cm}
  \epsfxsize=8cm
 \epsfysize=6cm
 \epsffile{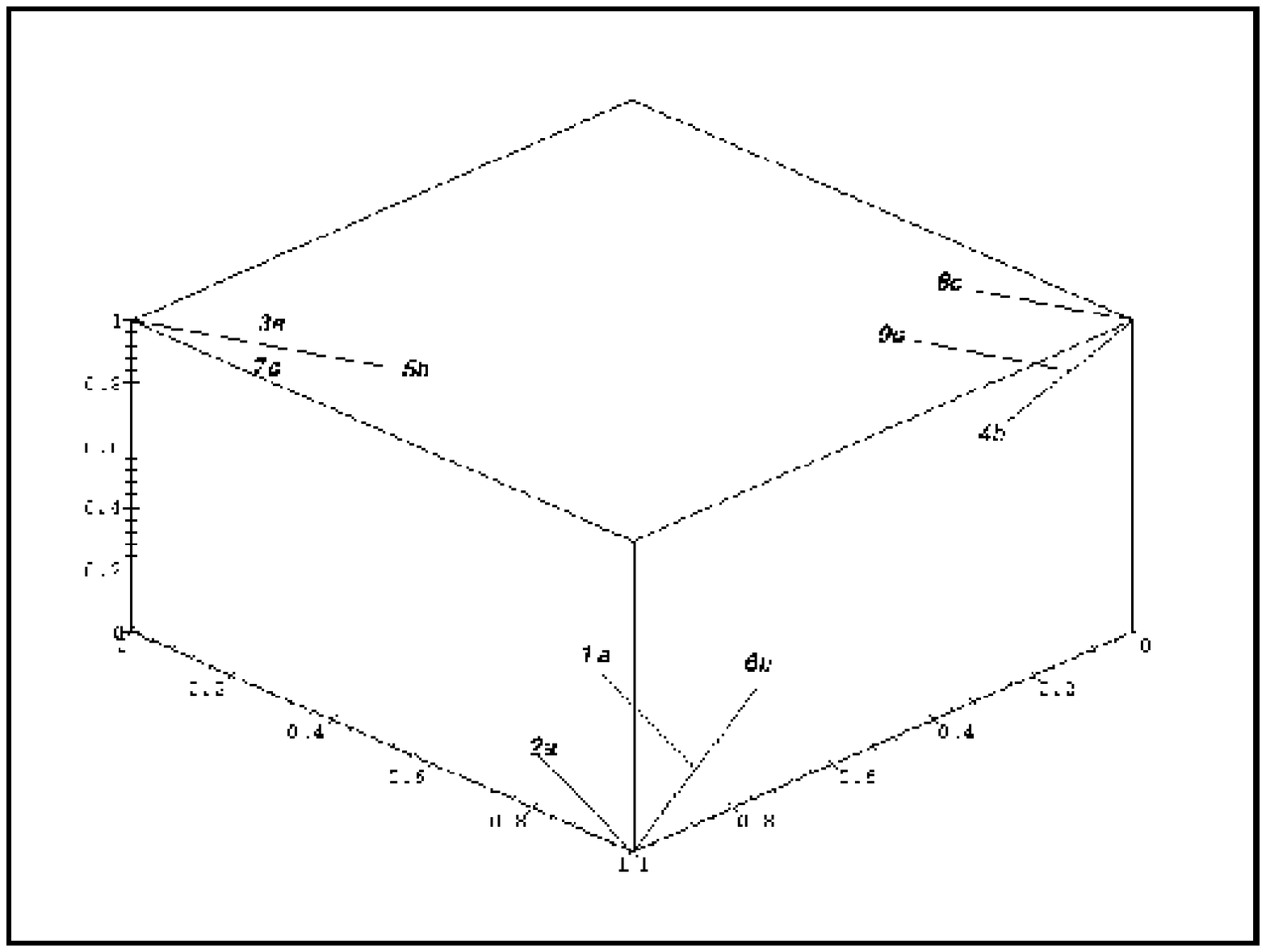}
\end{minipage}\\
\begin{minipage}{8cm}
  \epsfxsize=8cm
 \epsfysize=6cm
 \epsffile{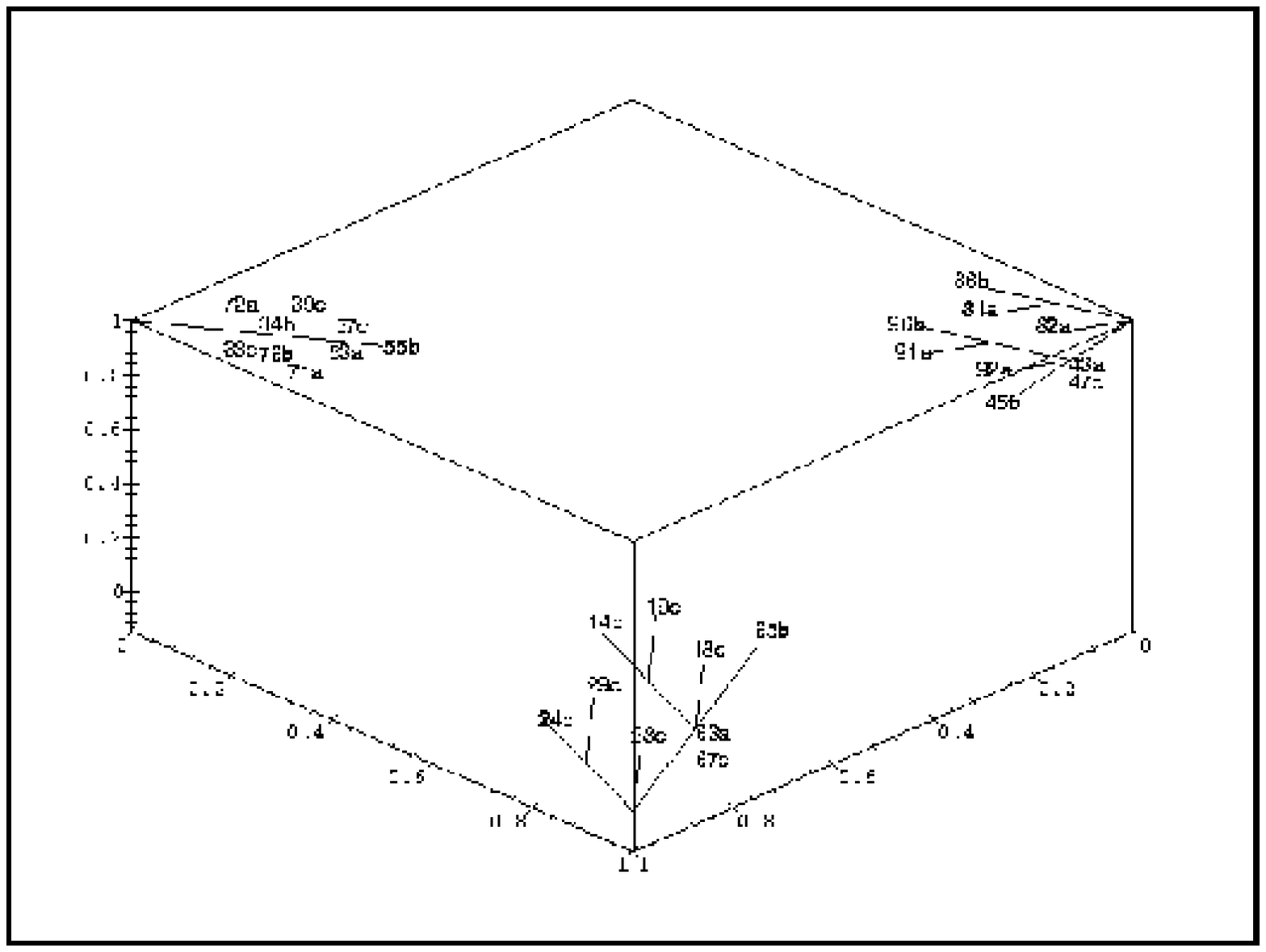}
\end{minipage}
\hspace{1cm}
\begin{minipage}{8cm}
  \epsfxsize=8cm
 \epsfysize=6cm
 \epsffile{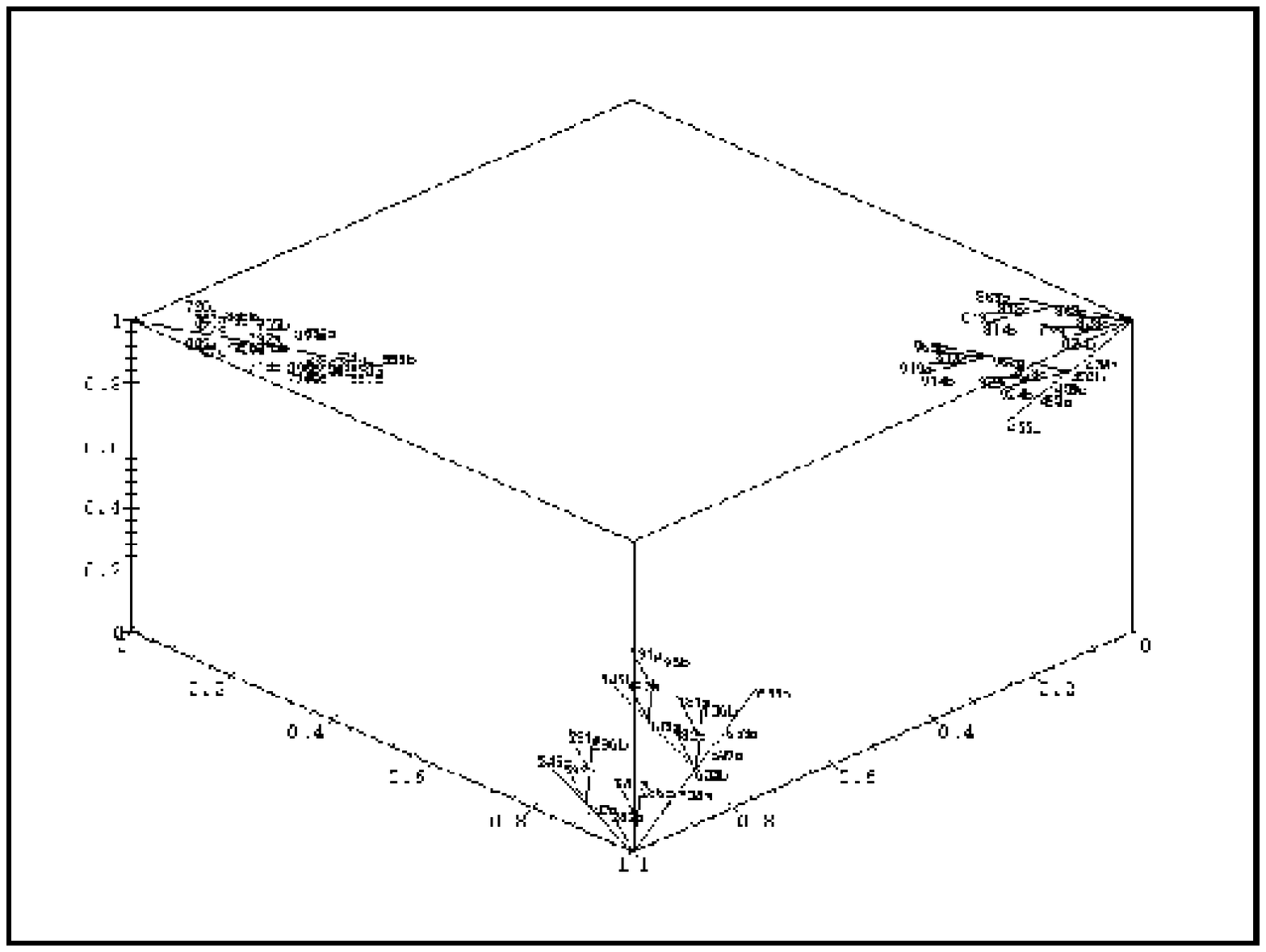}
\end{minipage}\\
\begin{minipage}{8cm}
  \epsfxsize=8cm
 \epsfysize=6cm
 \epsffile{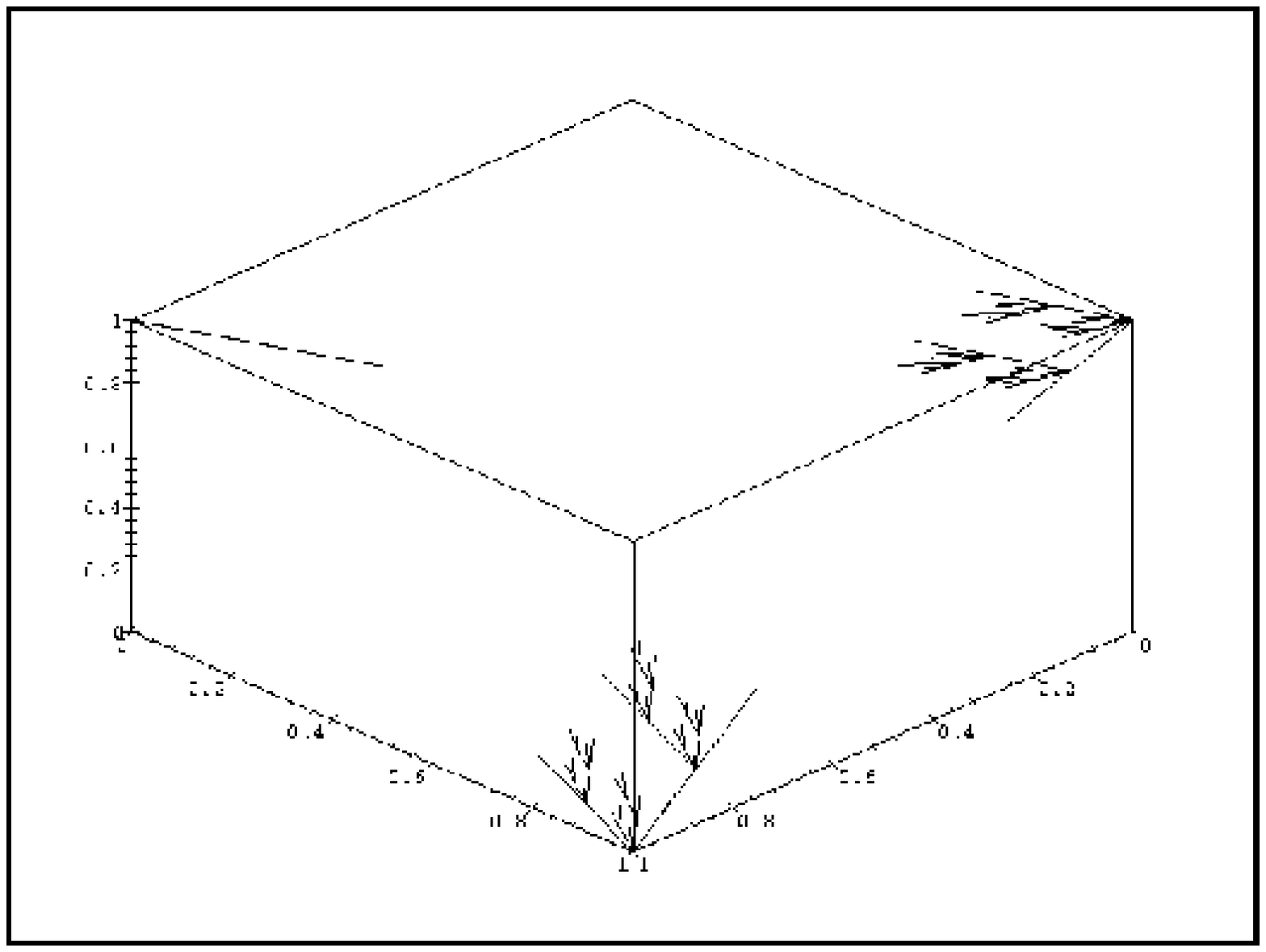}
\end{minipage}
\hspace{1cm}
\begin{minipage}{8cm}
  \epsfxsize=8cm
 \epsfysize=6cm
 \epsffile{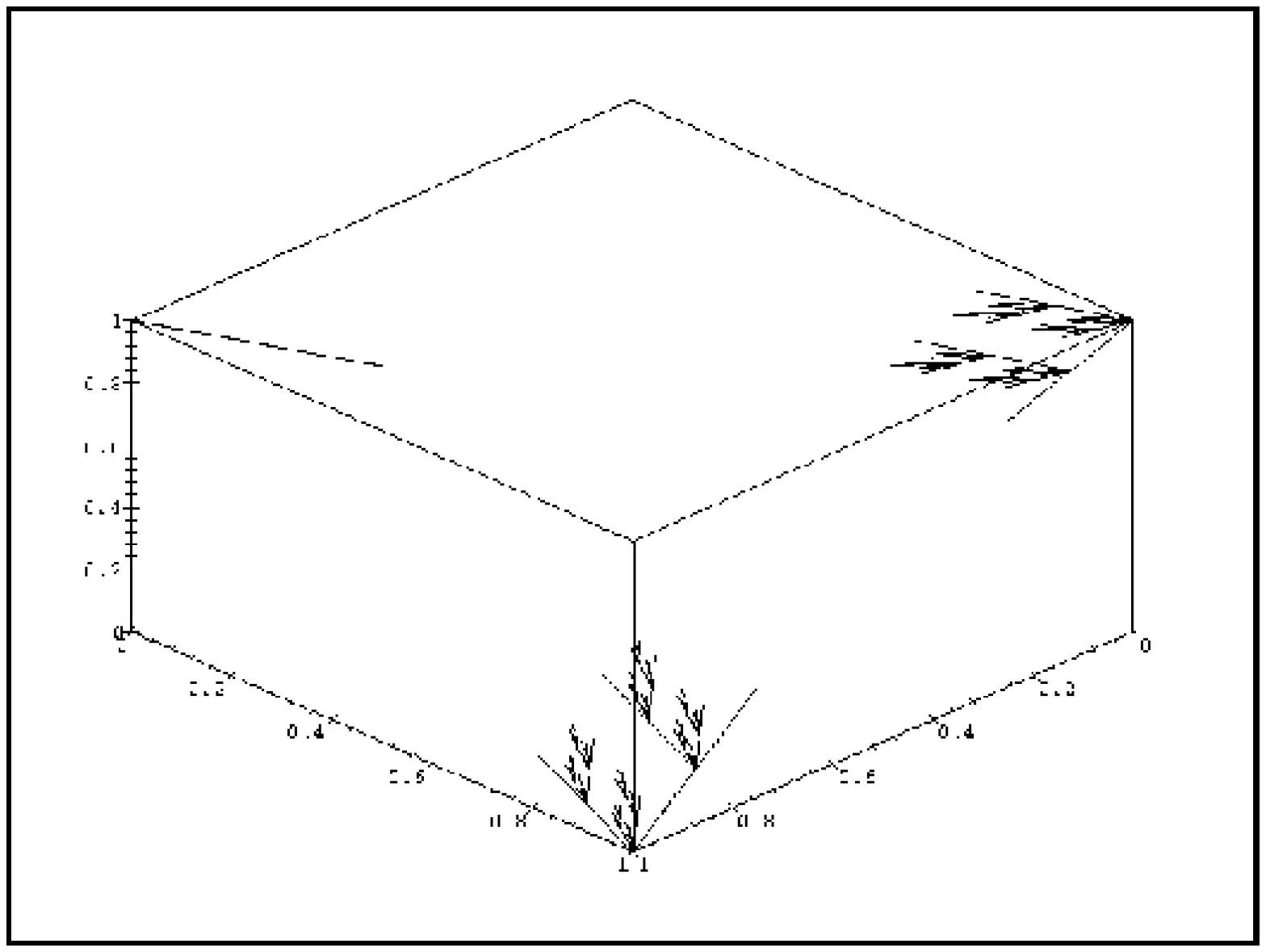}
\end{minipage}
\vspace{0.3cm}

\caption{ Five first steps of iteration of the Iterated Function System , for $E_c =1, \ N=3, \epsilon =0$.
 The labeling of the branches of the tree,
is a sequence, read from the left to the right, indicating the sequence of mappings applied to one of the initial branch $a, b, c$
(right most symbol). The last picture is the plot of a trajectory on the attractor.} \label{F12}
 \ec
 \enf

\sssu{ One dimensional case where $E_c \in
]\frac{1+\epsilon}{1-\epsilon},\frac{2}{1-\epsilon}]$.} \label{atomic}

This case is very atypical in the sense that the attractor is
a finite set of points whose components have values $\frac{1+\epsilon}{1-\epsilon}$.
Indeed, let ${\cal I}$ be the set where each site has energy
$X=\frac{1+\epsilon}{1-\epsilon}$ but at most one with energy $\frac{2.\epsilon}{1-\epsilon}$.
For $E_c \in ]\frac{1+\epsilon}{1-\epsilon},\frac{2}{1-\epsilon}]$ the set $\calI$ is the unique invariant set.
Moreover, $\forall \calB \in \calM, \
\calT^n(\calB) \longrightarrow \calI$, in the Haussdorf metric. Note that in this case we do not have
invertibility, but each point has exactly $N$ preimages.

This behaviour is somehow pathological as it exists only in this range of $E_c$ value. For higher dimensions,
we still do not know if there can be such an atomic invariant set.

\ssu{The probability distribution of avalanches size.}

In this part we derive a  relation linking the sum of Lyapunov exponents and the
probability of avalanche size.
Then, we relate the fractal structure of the attractor to the critical exponent values.
The basic ingredient is the Ledrappier-Young formula. However, as already said,
the existence of a kernel in the standard Zhang's model makes the analysis somehow cumbersome.
We therefore discuss first the non kernel case ($\epsilon \neq 0$) and comment only briefly on the modifications
necessary to handle the limiting case  $\epsilon = 0$.

\sssu{Average contraction rate.}

The key result is an exact formula linking the determinant of the basic maps to the total number $s$
of overcritical sites in the corresponding avalanche.
Namely we prove the following :

\bp
For $T^k_a \in K_s$ one has

\beq \label{det}
det L_a^k = \epsilon^s
\eeq
\ep

\bpr
We first show a propery about the relative distance of the overcritical sites in an 
avalanche. Let the avalanche be given by $A = \left\{A_k \right\}_{0 \leq k \leq n}$
where $A_k $ is the set of overcritical sites at the kth step in the avalanche $A$.
Denote by $D(A_k) = \left\{d(i,j) \ : \ i,j \in A_k  \right\}$ the set of pairwise 
distances of the vertex set $A_k$. 
The proof is a straightforward consequence of the following lemma.

\ble
For any $\bbbz^d$ sublattice $\Lambda$, $D(A_k) \subset 2.\bbbn \Rightarrow D(A_{k+1}) \subset 2.\bbbn$
\ele

Let $\gamma(i,j)$ denotes any path from $i$ to $j$ with no repetition of edges and 
$|\gamma |$ its length. From the general properties of subsets of $\bbbz^d$ it follows
that $d(i,j) \in 2.\bbbn \Rightarrow |\gamma(i,j)|  \in 2.\bbbn $ and, vice versa, if
$ |\gamma(i,j)|  \in 2.\bbbn$ for some $\gamma$ then $d(i,j) \in 2.\bbbn$. Let
$|A_k|,|A_{k+1}|  \geq 2 $ and $i,j \in A_k$. 
We will show below, that provided $\epsilon$ is sufficiently small, no site
can be overcritical for two successive time steps. Assuming this for the moment, it follows that
 $A_{k+1} \subset \calB( A_k,1)=
\left\{v \in \Lambda \ : \ d(v,A_k) =1 \right\}$ since no site of $A_k$ can be 
overcritical in the next step. Fix a path $\gamma^\ast$ by eliminating the first and last 
edge of $\gamma \subset A_k$.  $\gamma^\ast$ is a path between a vertex in $\calB(i,1)$ and $\calB(j,1)$,
 of length $|\gamma-2|$. Since the pairwise distance in $\calB(v,1)$  are even for any 
vertex $v \in \Lambda$ it follows that for any pair of vertices from 
$\calB(i,1)$ to $\calB(j,1)$ there is an even length extension of $\gamma^\ast$ connecting
those two vertices. This proves the lemma.\\

We now show that, provided $\epsilon$ is sufficiently small, a site cannot be overcritical
in two successive time steps. For $\epsilon=0$ this is obvious. Assume now
that $\epsilon>0$.  Let $\tilde{E}_k$ be the maximal energy value of an overcritical
site in the $k$th step of an avalanche. For a given $\epsilon$ we have to show that
$\epsilon.\tilde{E}_k < E_c, \forall k$. Clearly, $\tilde{E}_0< E_c+1$ . It is obvious that the
maximal increase of energy on a site $v$ can only happen if $v$ has $2.d$ overcritical neighbours.
In that case we have the following estimation:

$$\tilde{E}_{k+1} < (1-\epsilon).\tilde{E}_k + E_c$$

Iterating this expression we obtain:

$$\tilde{E}_n < (1-\epsilon)^n.\tilde{E}_0 + E_c.\sum_{i=0}^{n-1}(1-\epsilon)^i =
(1-\epsilon)^n.\tilde{E}_0 + Ec.\frac{1-(1-\epsilon)^n}{\epsilon}$$

\nid which has to be less than $\frac{E_c}{\epsilon}$. This holds provided :

\beq
E_c > \frac{\epsilon}{1-\epsilon}
\eeq

It follows from the lemma that two neighbours cannot be simultaneously overcritical 
during one avalanche provided $\epsilon$ 
sufficiently small. One then gets the expression for the determinant by decomposing the matrix of the avalanche
into one step matrices. The row corresponding to any overcritical site  as only one non zero 
entry, the diagonal element $\epsilon$ (nothing comes from the other overcritical sites at this 
time)  while the columns corresponding to a non overcritical site has only one non zero entry,
the diagonal element $1$. Formula (21) follows.
\epr

By using the ergodic theorem we get the log-average volume contraction which is also
the sum of Lyapunov exponents as:

\beq \label{Pl_Lyap}
\sum_i \lambda_i^- = log\epsilon . \sum_{s=1}^{S_N} s P_N(s) = log\epsilon.\bar{s}
\eeq

\nid where $\bar{s}$ is the average avalanche size and $s_N$ the maximal
avalanche size. The formula relates the {\it local volume
contraction} to the {\it average avalanche size}. It connects therefore microscopic dynamical quantities (Lyapunov exponents) to a macroscopic
observable (average avalanche size). In particular it allows to establish a link between
the Lyapunov spectrum and the critical exponents of the avalanche size distribution (see below
and \cite{BCK3})

\sssu{Contraction versus expansion.} \label{contrac_expan}

The average contraction rate
decreases with increasing $E_c$. Indeed, the larger $E_c$, the larger is the frequency of occurence of
"trivial" avalanches where no
relaxation occurs. They only display neutral directions in the phase space, with no contraction, and
no contribution to the
negative Lyapunov exponent. This can also be seen on formula (\ref{det}) : the larger $E_c$,
the smaller the average avalanche size.
 Therefore, for fixed $N$, there exists an $E_c^*(N)$ which is the unique $E_c$ value
such that:

\beq \label{Ec*}
log\epsilon.\bar{s} + log(N) = 0
\eeq

For $E_c < E_c^*(N)$ the contraction dominates the expansion, while it is the opposite for
$E_c > E_c^*(N)$. Clearly, the invariant set structure is  different
 in these two cases. On the one hand, for small $E_c$ values, the images of the domains $\calP_a^k$ are thin bands which
are stretched slower than they contract. Therefore, they are expected not to overlap asymptotically
and  the invariant set has a Cantor structure with large gaps. On the other hand, when $E_c > E_c^*(N)$, the successive images of the domains $\calP_a^k$  fill more
and more  the phase space and the properties in conjecture 1 and 4 should not hold. 

Note that, in this scheme, the Haussdorf dimension of $\calA$ 
increases for increasing $E_c$, $E_c < E_c^*(N)$ and is likely to be constant when
$E_c > E_c^*(N)$ .\\

The graph of $E_c^*(N)$ can easily be computed numerically. We give an example below, in a square lattice,  for various
values of $\epsilon$ (Fig. \ref{F_EcN}). Note that $E_c^*(N)$ increases with $N$. Therefore, one expects that,
conjecture \ref{IFS1} \ref{Invertibility}  hold on larger and larger range of
$E_c$ values, as $N$ increases.  

\bef
 \bc
 \begin{minipage}{8cm}
  \epsfxsize=8cm
 \epsfysize=6cm
 \epsffile{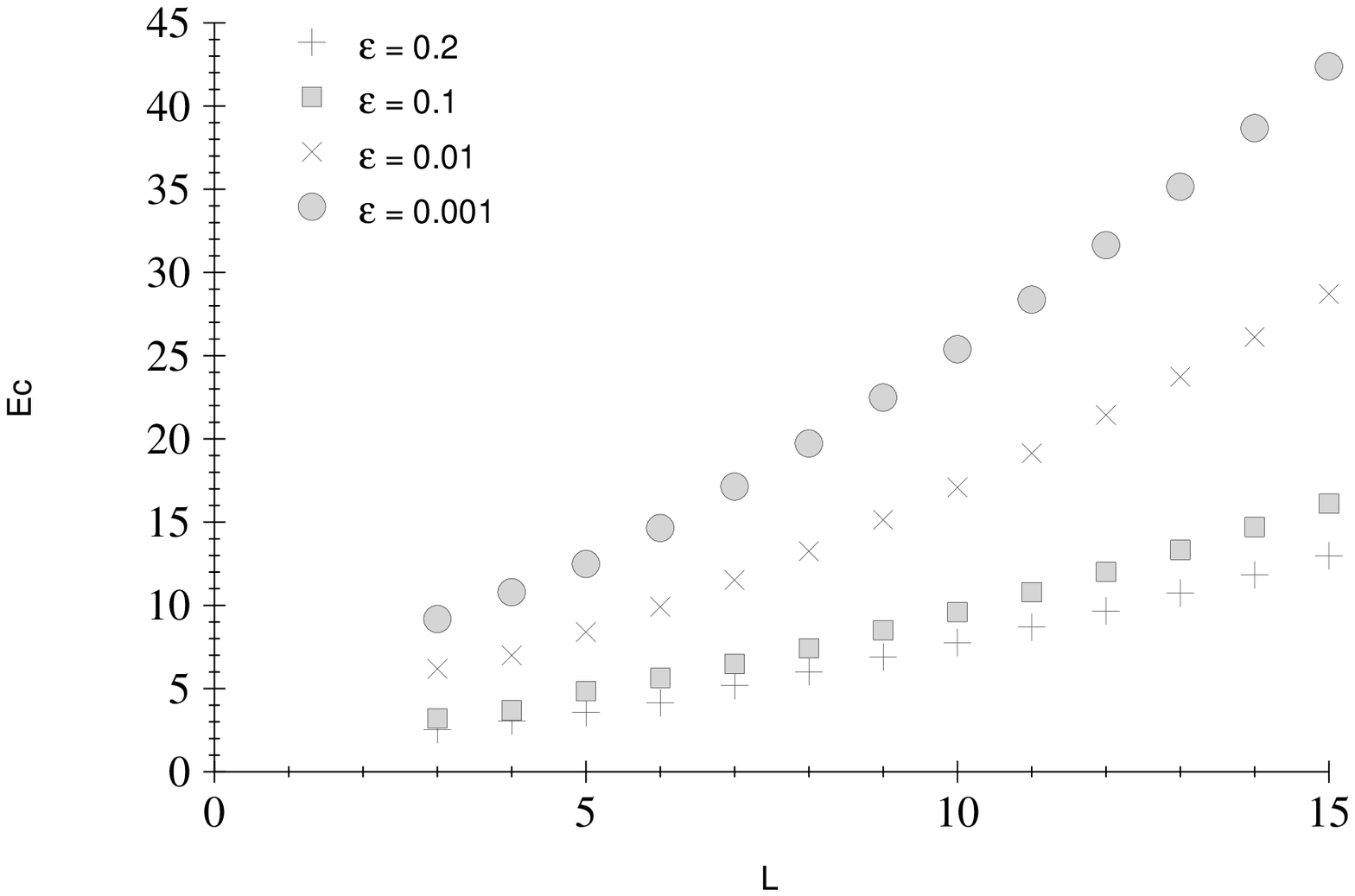}
\end{minipage}
\hspace{1cm}
\vspace{0.3cm}

\caption{$E_c^*(N)$ for various values of $\epsilon$.
$L$ is here the linear dimension of $\Lambda$ ($N=L^2$). \label{F_EcN}
}
 \ec
 \enf

\sssu{Bounds on the critical exponent.}

In the invertible case the Ledrapier-Young formula implies:

\beq
\sum_{i=1}^N |\lambda_i | \geq log N
\eeq

Therefore, from formula (\ref{Pl_Lyap}) :

\beq \label{s}
 \sum_{s=1}^{s_N} s   P_N(s) \geq \frac{logN}{|log\epsilon|}
\eeq

This implies that the average avalanche size, $\bar{s}$ , has to diverge when
$N$ goes to infinity and that,
in the thermodynamic limit, $P_N(s)$ tends to a distribution with an infinite mean-value.

Furthermore, a reasonable assumption (supported by experiments) is that, for fixed $N$,
the probability decreases with the avalanche size, namely:

\beq \label{Decroissance}
\forall N \geq N(\epsilon), \ P_N(s) \geq P_{N}(s+1)
\eeq

In a certain way, this behaviour could be expected since the larger the avalanche, the more one
has to impose conditions defining the corresponding domain of continuity, and the less the corresponding
volume. However, this argument is not completely correct in general since one assumes some kind of absolute
continuity of the invariant measure on the stable foliation (the probability of a domain
decreases with its volume). In particular it is completely false for $Ec > 1$ in the one dimensional chain,
here the probability {\it increases} with $s$.

Assuming that that there is indeed a power law, and that the system is invertible 
then one obtains:

\beq \label{scaling}
P_N(s) = \frac{f_N(s)}{s^{\tau}}, \ 1 < \tau \leq 2 
\eeq

\nid where $f_N(s)$ is a cut-off function accounting for finite size effects 
\footnote{In the SOC litterature, the 
Finite Scaling Assumption leads to write the probability distribution of avalanche size 
$P_N(s)$ as $P_N(s)= s^{-\tau} \calG(s.L^{-\beta})$  where 
$\calG$ accounts for cut-off effects. The exponents $(\tau,\beta)$ are believed to characterize
the universality class of the model. Note that $\tau,\beta$ depend a priori from $Ec$.}.

Therefore, eq. (26) gives {\it the scaling of the power law and bounds for
the critical exponent $\tau$}. In particular, if we assume that $P_N(s)$ converges to some
limit $P^\ast(s)$ as $N\to + \infty$, then $P^\ast(s) = \frac{c}{s^{\tau}}, \ \tau \in [1,2]$.

\sssu{The value of $\tau$ and the fractality of the support of the invariant measure.}

The Ledrappier-Young formula gives a direct way to check the ``fractality'' of the support
of the invariant measure in the invertible case. For

\beq
|\sum_i \lambda_i^-| > log N
\eeq

\nid if some partial dimensions $\sigma_i$ are not integers ($< m_i$).

Suppose that the measure $\mu$ is absolutely continuous on the stable foliation.
Then from eq. (\ref{LY}) we get equality in (\ref{s}), implying that the sum
diverges logarithmically with $N$. Furthermore,  the maximal avalanche size 
scales like $s_N \approx N^{\beta/2}$, implying
that $logs_N \approx logN$. Then $\bar{s}$ diverges {\it logarithmically with $s_N$ 
suggesting a critical exponent $\tau = 2$}.
More generally, we get the same result if the fractal set is homogeneous (in
sense that all partial Haussdorf dimensions are equal or can be bounded from
below as $N \longrightarrow \infty$).

However, one does not expect the fractal to be homogeneous.
 It is indeed clear that the contraction is not uniform in the phase space.
For $\epsilon=0$ the kernel directions produce infinite contraction. In figure \ref{F12} they are the directions
transverse to the "branches" of the attractor, which project the dynamics on the tree, in one time step.
On the other
hand, the directions "parallel" to the branches produce finite contraction. As a corrolary, the partial Hausdorff
 dimensions of the invariant set
are zero transversally to the attractor  while they are finite along the branches.
When $\epsilon$ is small, there are still directions
producing high contractions, those which give the kernel directions as $\epsilon  \rightarrow 0$.
This effect is reflected in the Lyapunov spectrum where one detects two parts in the spectrum (see Figure \ref{F_Lyap}).

\bef
 \bc
 \begin{minipage}{8cm}
  \epsfxsize=8cm
 \epsfysize=6cm
 \epsffile{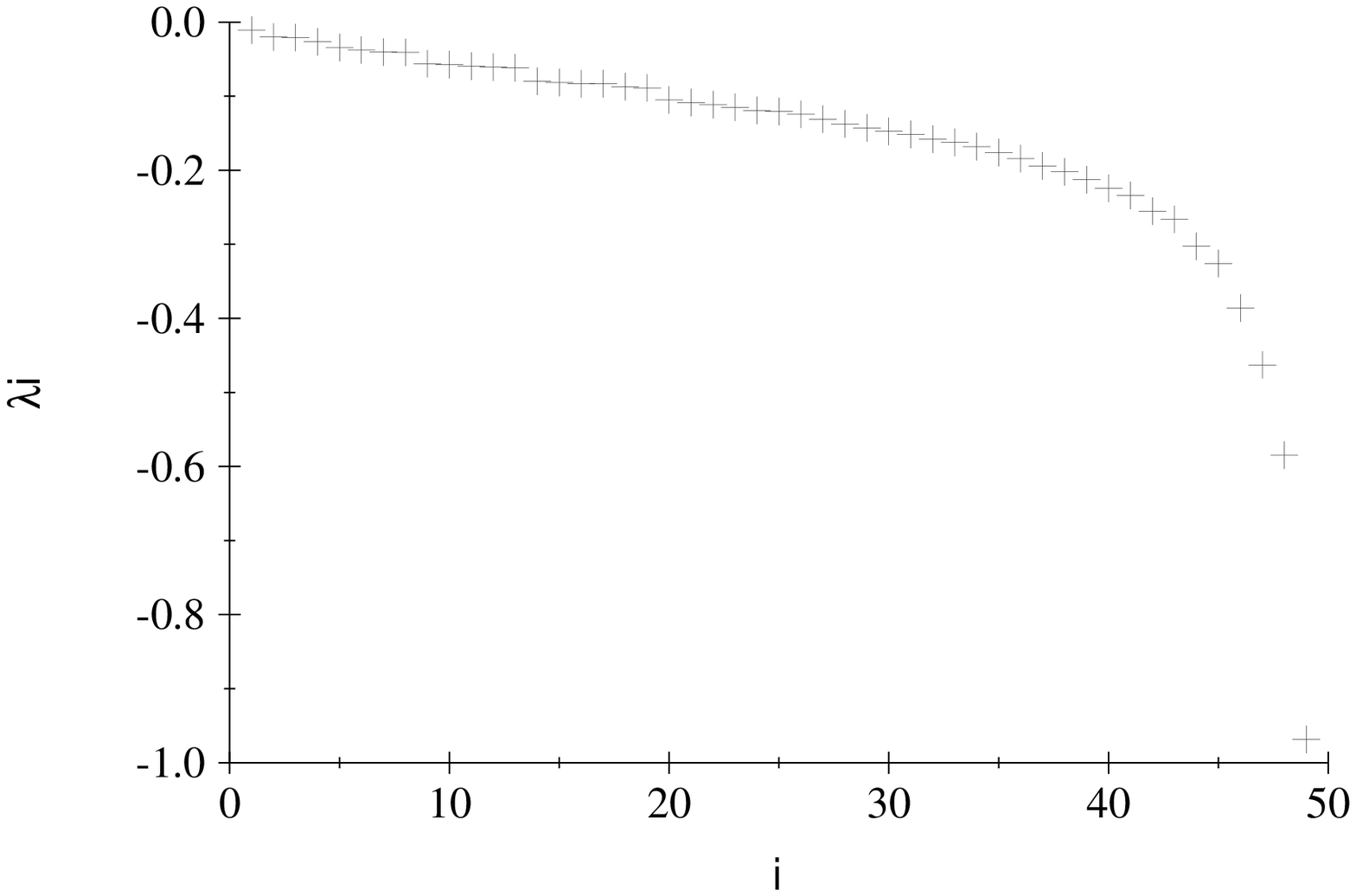}
\end{minipage}
\hspace{1cm}
\vspace{0.3cm}

\caption{Spectrum of negative Lyapunov exponents for $N=49, \ Ec= 3, \ \epsilon =0.1$. \label{F_Lyap} }

 \ec

 \enf

Therefore, the sum (resp. $\bar{s}$)  must diverge faster than logarithmically with $N$  and  strict
inequality holds. This  implies that the critical exponent $\tau < 2$.
The measured exponent is indeed always strictly lower than 2 \cite{Luebeck}. Note that under the finite size scaling hypothesis $\bar{s}$ behaves as $N^{1-\frac{\tau}{2}}$ and 
 that $\tau$ is indeed lower than 2 iff strict inequality holds  
in (\ref{s}). \\

An explicit formula linking the Hausdorff dimension and the critical exponents
 $(\tau,\beta)$ can be obtained through
the Ledrappier-Young formula. This will be treated in
a separated paper \cite{BCK3}.

\sssu{The case $\epsilon = 0$}

One would like to obtain an equality like (\ref{s}) also in this case.
However, setting $\epsilon = 0$ leads  to a zero determinant an
hence infinite Lyapunov exponents (more precisely the Lyapunov
exponents are not defined on the whole configuration space). 
But restricted to $\calE^s(\hat{\bX})$ one can still compute finite Lyapunov exponents and the sum is just the determinant of the matrix {\it
restricted} to the stable space $\calE^s(\hat{\bX})$ at some point
$\hat{\bX}$ in the domain of the map.

If one has a furhter inequality like:

\beq
det(T_a^k |_{\calE^s(\hat{\bX})}) \geq C^s
\eeq

\nid $\forall K \in K_s$ and $0<C<1$ one still gets the same kind of estimates for the expected avalanche size like in the $\epsilon \not= 0$ case. The details are quite cumbersome and will be given in a forthcoming paper.

\ssu{Phase transitions.} \label{Phase_Trans}

The domains of continuity $\calM_a^k$ are bounded by
hyperplanes, {\it which are moving when $E_c$ varies}. In general, a small variation in $E_c$
does not lead to structural changes in the dynamics, if all these hyperplanes are intersecting 
the interior of $\calM$.
In this case, the structure of the transition graph is not modified.
Moreover, the corresponding mapping $T_a^k$ does not change under this motion.
More precisely, changes in $E_c$ just
change the shape of $\calM_a^k$ but not the matrix of the mapping $T_a^k$.

However, for some $E_c$ values, some hyperplanes {\it
have intersection only with $^\partial\calM$}. This implies that a small change in $E_c$ can 
push these hyperplanes outside $\calM$. Hence the corresponding transition graph changes in structure.
As far as the asymptotic dynamics and therefore, the invariant distribution is dependent
on the graph structure, we expect changes in the SOC picture when crossing
these {\it critical } $E_c$ values.
This effect has already been reported elsewhere for the one dimensional
Zhang's model \cite{BCK1} and arises also in two dimensions where $P_N(s)$
is not a power law for $E_c << 1$ \cite{BCK3}.

In fact, one can easily figure out that at least
the limiting cases $E_c \rightarrow \infty $ and $E_c \rightarrow 0 $ are completely different.
For $E_c \rightarrow \infty $ relaxation events are more and more seldom.
 One obtains kind of a frozen state where energy increases (on average)
monotonously with some rare (but large) avalanches.  Moreover, the asymptotic energy distribution is 
sensitive to the initial conditions (loss of ergodicity). Furthermore, the
attractor as a large Haussdorf dimension.

 On the other hand, for $E_c \rightarrow 0 $, each activation  generates a very large avalanche (that has to reflect
many times on the boundary before it has lost  enough energy to stop). 
This implies larger and larger contraction, and therefore the sum of Lyapunov exponents decreases to 
$- \infty$. As a corollary  of Ledrappier-Young formula 
the partial fractal dimensions have to go to zero in order to maintain
the product equal to $logN$.

\su{Conclusion.}

We have shown that certain classes of models of SOC like the Zhang's model fit
naturally into a well known class of dynamical systems. Especially for  the
question of asymptotic energy distribution, observables distribution,
ergodicity,  this seems to be a proper point of view.  Furthermore it seems
likely to exhibit close relationship between the probability  of the size of
avalanches and the fractality of the attractor. \\

There are many questions for further investigations. We list a few of them.

\ben

\item {\bf Development of a thermodynamic formalism and it's linkage to the SOC quantities}.
It should be possible to extrapolate this formalism to the case of arbitrary (hyperbolic) SOC-system. Moreover,
 phase transitions 
 should correspond to changes in the invariant
measure of maximal entropy (loss of analyticity of 
the topological presssure). One expects that in a proper formulation $E_c$
should play the role of an inverse temperature.

\item {\bf Dimension spectrum of the attractor and Lyapunov spectrum}. We have outlined above the 
crucial role played by Lyapunov exponents (accounting for energy transport) and the link one can establish with the equilibrium state and 
the critical exponents. The full developments of this point will be published elsewhere \cite{BCK3}

\item {\bf Nonuniform distribution rules.} As outlined in the paper most of
our results carry on if one does not choose a uniform activation measure, because
one still has a good measure as an equilibrium state. On the other hand, activating with a degenerate
probability distribution (for example activating always the same site) will lead to different results. Activating 
sites periodically with different period will allow to sample the periodic orbits structure 
of the global attractor, which are dense.

\item {\bf Thermodynamic limit for fixed $E_c$ and $N\to\infty$ and
 the limit $E_c \to \infty$ ($N$ fixed).} In these both cases one loses the
hyperbolic structure.

\item{\bf Smooth thresholds.} Some modification of Zhang's model have been proposed, in particular to treat this model in the continuum
limit by an anomalous diffusion equation \cite{Diaz,Bantay}. In this case the Heaviside function corresponding to the sharp threshold at 
$E_c$ is smoothed out by some continuous function. The nice effect of this change in our description is that it removes the singularity set.
 On the other hand, the system is expected to have still a nice hyperbolic structure (though non uniform) where smooth local stable manifolds 
exist for almost all points.  Pesin theory \cite{Pesin}
should apply in this context.

\item{\bf The case where $\delta X$ is random.} In the usual Zhang's model, the energy activation quantum $\delta X$ is not a constant
but is a random
variable. This situation can  be treated in the framework of random hyperbolic dynamical system.

\een

As a conclusion we would like to outline  that the study of SOC-models with 
tools from dynamical system theory will certainly {\it not solve} all
questions in this context. In the belief of the authors it is mainly useful for the study of fairly
 general structure properties of the models. It is also clear that the complexity of the underlying
transition graph on which the model is defined will become of crucial importance for some questions. \\

{\bf Acknowledgement.}
This work has been partially supported by the PROCOPE grant (N 98008). 
B.C. is very greatfull to D. Searles
for helpful discussions.

\pagebreak
\listoffigures
\ed